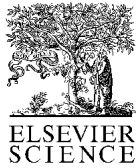 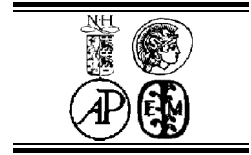

# Hydrostatic pressure study of the structural phase transitions and superconductivity in single crystals of $(Ba_{1-x}K_x)Fe_2As_2$ ($x$ = 0 and 0.45) and $CaFe_2As_2$

M. S. Torikachvili,[a,*] S. L. Bud'ko,[b] Ni Ni,[b] and P. C. Canfield[b]

[a]Department of Physics, San Diego State University, San Diego, CA 92182-1233, USA

[b]Ames Laboratory and Department of Physics and Astronomy, Iowa State University, Ames, IA 50011, USA

**Abstract**

We studied the effect of hydrostatic pressure ($P$) on the structural phase transitions and superconductivity in the ternary and pseudo-ternary iron arsenides $CaFe_2As_2$, $BaFe_2As_2$, and $(Ba_{0.55}K_{0.45})Fe_2As_2$, by means of measurements of electrical resistivity ($\rho$) in the 1.8 – 300 K temperature ($T$) range, pressures up to 20 kbar, and magnetic fields up to 9 T. $CaFe_2As_2$ and $BaFe_2As_2$ (lightly doped with Sn) display structural phase transitions near 170 K and 85 K, respectively, and do not exhibit superconductivity in ambient pressure, while K-doped $(Ba_{0.55}K_{0.45})Fe_2As_2$ is superconducting for $T$ < 30 K. The effect of pressure on $BaFe_2As_2$ is to shift the onset of the crystallographic transformation down in temperature at the rate of ~ –1.04 K/kbar, while shifting the whole $\rho(T)$ curves downward, whereas its effect on superconducting $(Ba_{0.55}K_{0.45})Fe_2As_2$ is to shift the onset of superconductivity to lower temperatures at the rate of ~ –0.21 K/kbar. The effect of pressure on $CaFe_2As_2$ is first to suppress the crystallographic transformation and induce superconductivity with onset near 12 K very rapidly, i.e., for $P$ < 5 kbar. However, higher pressures bring about another phase transformation characterized by reduced resistivity, and the suppression of superconductivity, confining superconductivity to a narrow pressure dome centered near 5 kbar. Upper critical field ($H_{c2}$) data in $(Ba_{0.55}K_{0.45})Fe_2As_2$ and $CaFe_2As_2$ are discussed. © 2001 Elsevier Science. All rights reserved

## 1. Introduction

The recent discovery of superconductivity (SC) at moderately high temperatures (T) in iron arsenide compounds with general compositions $R$FeAsO ($R$ = rare earth) and $A$Fe$_2$As$_2$ ($A$ = alkaline earth metal) triggered great interest in these materials.[1-5] Although the parent compounds show spin, charge, or crystallographic phase transitions in the 80 – 200 K T-range, and are not SC at ambient pressure (P), SC can be realized upon doping (chemical pressure effects are implicit),[1-5] or by applying pressure.[6-8]

---

* Corresponding author. e-mail: milton@sciences.sdsu.edu.

The $A$Fe$_2$As$_2$ compounds ($A$ = Ca, Ba, and Sr) form with the ThCr$_2$Si$_2$-type tetragonal structure (space group I4/mmm). Crystallographic phase transformations from the high-T tetragonal to the low-T orthorhombic phase take place near 170 K,[9] 140 K,[10] and 205 K,[11] respectively. Although these materials are not SC at ambient pressure, SC can be realized upon doping at the $A$- and Fe-sites,[4, 12-14] as well as upon the application of pressure;[6-8] in both cases SC occurs only when the phase transformation from tetragonal to orthorhombic structure is suppressed.

Whereas chemical substitution is an effective method for altering the density of states at the Fermi level, the lattice parameters, and perhaps the stability of a crystallographic phase, it introduces disorder that can affect the physics in a number of uncontrollable ways. On the other hand, hydrostatic pressure is a much less disruptive tool for probing the electronic properties, and its relationship to structural transformations. In this work we describe a systematic study of the effect of hydrostatic pressures up to 20 kbar in the electrical resistivity of pure CaFe$_2$As$_2$ and BaFe$_2$As$_2$, and K-doped Ba$_{0.55}$K$_{0.45}$Fe$_2$As$_2$.

## 2. Experimental Details

Single-crystals of CaFe$_2$As$_2$ and Ba$_{1-x}$K$_x$Fe$_2$As$_2$ ($x$ = 0, 0.45) were grown from Sn-flux, using conventional flux-growth techniques.[15, 16] The crystals grow as small platelets with facets perpendicular to the c-axis; they are quite fragile, and exfoliate easily. Most crystals have droplets of Sn on the surface. A crystallographic analysis suggests that the Ba-containing crystals may have incorporated a small amount of Sn in the structure.[16] The measurements of electrical resistivity under pressure were carried out with a self-contained piston-cylinder type Be-Cu pressure cell, with a core of hardened NiCrAl alloy (40KhNYu-VI). The sample, manganin, and Pb manometers were mounted on a feedthrough, which was inserted into a teflon capsule filled with either Fluorinert FC-75 or a 60:40 mixture of n-pentane:light mineral oil, which served as pressure transmitting media. Pressure was generated at room temperature with an hydraulic press, using manganin as a reference manometer. The pressure was locked in, and the cell was then loaded into a Quantum Design Physical Property Measurement System (PPMS-9), which provided the temperature and magnetic field (H) environments for these measurements, as well as the dc measurements of resistance for the sample (I//ab-plane), manometers, and a cernox temperature sensor attached to the body of the cell. The pressure at low temperatures was determined from the SC $T_c$ of pure Pb. The cooling and warming rates were kept below 0.5 K/min; the T-lag between the cernox sensor and the sample was < 0.5 K at high-T, and < 0.1 K below ~ 70 K. In light of the nearly linear variation of pressure from room temperature to ~ 90 K in piston-cylinder cells,[17] the pressure values at temperatures between these limits were estimated from a linear interpolation.

## 3. Experimental Results

CaFe$_2$As$_2$ displays a first order crystallographic transformation from high-T tetragonal to low-T orthorhombic near $T_{s1}$ ~ 170 K, and it is not superconducting at ambient pressure.[9] However, the application of modest hydrostatic pressures ($P$ < 5 kbar) suppresses the crystallographic transformation at $T_{s1}$ and drives it to lower temperatures at the very fast rate of ~ –12 K/kbar. The $\rho(T)$ data of Fig. 1 (I//ab-plane) show that superconductivity with onset near 12 K is stabilized in pressures between 2.3 and 8.6 kbar, when the transition at $T_{s1}$, as identified by its feature in $\rho(T)$, is either partially or completely suppressed. The onset of SC is accompanied by a drop in residual resistivity just above $T_c$ as a function of pressure, as indicated in Fig.1b. The $\rho(T)$ curves at 5.5, 12.7 and 16.8 kbar show clearly that a second phase transition takes place at $T_{s2}$ ~ 102, 155 and 235 K (as identified from maxima in $d\rho/dT$), respectively, and that this high-P/low-T phase has reduced scattering. For simplicity, the only data shown in Figs. 1a and 1b are from the cooling cycle; the $\rho(T)$ data taken upon warming (data not shown) show significant hysteresis both at $T_{s1}$ and $T_{s2}$, suggesting that these are both first order transitions. In light of the neutron scattering results, indicating that a structural transition from a high-P/high-T tetragonal to a high-P/low-T collapsed-tetragonal phase at 6.3 kbar, takes place between 100 and 150 K (the change is cell volume is ~ 5%),[18] it is tempting to ascribe the break in $\rho(T)$, separating the higher and lower resistivity regions, to this second structural transition. $T_{s2}$ increases very rapidly with pressure at the rate of ~ 15 K/kbar, while SC is suppressed in the collapsed phase. Therefore, superconductivity is confined to the range in which the pressure is high enough to suppress the tetragonal-to-orthorhombic structural transformation at $T_{s1}$, but lower than necessary to fully stabilize the reduced-resistivity collapsed-tetragonal phase at $T_{s2}$. The pressure-temperature phase diagram of Fig. 1c displays $T_{s1}$, $T_{s2}$, and $T_c$ as a function of $P$; shown as well are the values of the residual resistivity just above $T_c$, at 15 K. This residual resistivity $\rho_{15K}$ is perhaps a gauge for the stability of the reduced-resistivity phase, which precludes superconductivity.

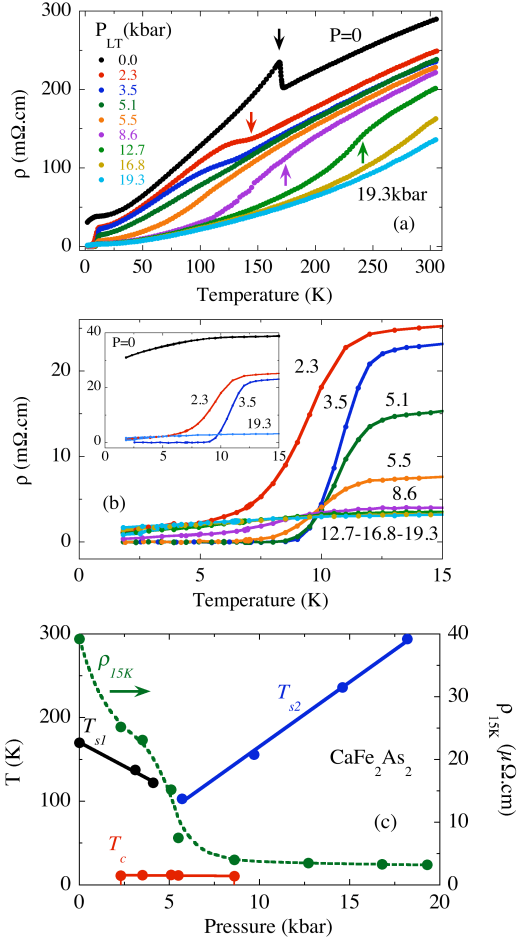

Fig. 1 – (color online) (a) $\rho(T)$ (I//ab) for CaFe$_2$As$_2$ in various pressures below 20 kbar. The down arrows indicate the onset of the low-P structural transformation for 2 pressures, whereas the up arrows indicate the onset of the high-P-reduced-resistivity phase; (b) zoomed-in view of $\rho(T)$ at low-T, detailing the evolution from non-SC behavior at $P$ = 0, to incomplete SC transition at 2.3 kbar, to complete SC transitions, incomplete SC transitions, and normal behavior at higher pressures; (c) pressure-temperature phase diagram delineating the phases determined by $T_{s1}$, $T_{s2}$, and $T_c$; it also shows the pressure dependence of the residual resistivity at 15 K. The lines are guides to the eye.

In order to characterize the pressure-induced SC state, we carried out measurements of $\rho(T)$ in fields up to 9 T, and determined the upper critical field ($H_{c2}$). The values of $T_{c,onset}$ (heretofore referred to as $T_c$) were taken from a 10% drop of the normal state resistivity preceding the onset of superconductivity.

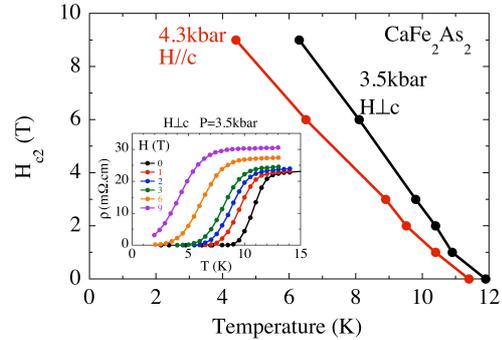

Fig. 2 – (color online) Upper critical field $H_{c2}$ vs $T$ for CaFe$_2$As$_2$ in $P$ = 3.5 kbar and 4.3 kbar, for H perpendicular and parallel to the c-axis, respectively. The values of $T_c$ were taken from a 10% drop of the normal state resistivity values preceding SC. The inset shows $\rho_{3.5kbar}(T)$ curves in fields up to 9 T, for H⊥c.

The behavior of $H_{c2}$ vs $T$ for CaFe$_2$As$_2$ at 3.5 kbar (H$\perp$c) and 4.3 kbar (H//c) are shown in Fig. 2. These $H_{c2}$ data are noticeably anisotropic. A naive extrapolation to $T = 0$ suggests $H_{c2}(0)$ values close to 15 and 20 T for H//c and H$\perp$c, respectively. As shown in the inset of Fig. 2, the field broadens slightly the SC transition from ~ 2.5 K at H = 0 to ~ 4.0 K at 9 T. It should also be noted that the magnetoresistance just above $T_c$ is positive and very large, reaching about 33% at 9 T.

Although the $\rho(T)$ data in CaFe$_2$As$_2$ under pressure, and the behavior of $H_{c2}(T)$ in fields up to 9 T are quite compelling evidences for bulk superconductivity, measurements of magnetization or specific heat under pressure are in order for a more definite conclusion.

The temperature dependence of the electrical resistivity $\rho(T)$ measured along the a-axis of single-crystalline BaFe$_2$As$_2$ in pressures up to 20 kbar is shown in Fig. 3. This specimen is lightly doped with Sn, an unintended consequence of the crystal growth process. The doping level is smaller than 1 at%, which is large enough for detection by wavelength dispersive x-ray spectroscopy, and by its effect on the lattice parameters, but too small for a reliable determination of site occupancy.[16] The behavior of $\rho(T)$ at ambient pressure for the lightly Sn-doped sample of this study is quite different from the behavior in polycrystalline specimens,[10] or single-crystals grown out of Fe-As self-flux,[19] possibly reflecting the incorporation of Sn as a dopant. $\rho(T)$ drops sharply below ~ 140 K both in Fe-As grown single crystals and polycrystalline samples,[10, 19] while $\rho(T)$ for the Sn-doped specimen of this study shows a broad minimum near $T_{min}$ = 135 K, precursive to the onset of the structural transition from high-T tetragonal to low-T orthorhombic. Taking the temperature of the structural transition from the point of highest inflection (minimum in $d\rho/dT$) yields $T_s$ ~ 85 K at P = 0, consistently with the value yielded by X-ray diffraction measurement.[16] In contrast to CaFe$_2$As$_2$, the behavior of $\rho(T)$ in BaFe$_2$As$_2$ upon cooling or warming doesn't show any noticeable hysteresis. A second feature in $\rho(T)$ near 50 K can also be observed. The main effects of pressure are 1) to gradually suppress the feature below 50 K; 2) to shift down the $\rho(T)$ curves for $P > 10$ kbar; and 3) to shift down in temperature both $T_{min}$ at the initial rate $dT_{min}/dP$ ~ –2.4 K/kbar, and $T_s$ at the rate $dT_s/dP$ ~ –1.0 K/kbar. At these rates, the structural transformation would be suppressed in the 55-80 kbar range. Actually, superconductivity in BaFe$_2$As$_2$ under pressure has been reported, for the P-range between 25 and 60 kbar.[8]

The temperature dependence of the normalized electrical resistivity of superconducting (Ba$_{0.55}$K$_{0.45}$)Fe$_2$As$_2$ in various pressures is shown in Fig. 4. The absolute value of $\rho_{300K}$ varied with pressure in ways that suggest that pressure improves the mechanical/electrical contact between the layers. However, the normalized resistivity $\rho/\rho_{300K}$ is nearly identical for all pressures down to temperatures close to $T_c$ (data not shown). The normal-state resistivity just above the onset of superconductivity drops with pressure, and the value of $T_{c, onset}$ drops at the rate of ~ –0.21 K/kbar. The SC transition broadens noticeably with pressure. Taking the outset of the SC transition at the 90% drop from the normal-state $\rho$, $T_{c, outset}$ drops at the slightly higher rate of ~ –0.25 K/kbar. The basal plane $\rho(T)$ data for (Ba$_{0.55}$K$_{0.45}$)Fe$_2$As$_2$ for $P$ = 0 and 20.4 kbar, in magnetic fields up to 9 T are shown in Fig. 5. The magnetoresistance just above $T_c$ is positive and moderately large. The broadening of the $\rho(T)$ curves with field increases noticeably under pressure. Neglecting the low field upturns of the $H_{c2}(T)$ curves (Fig. 5 inset), the values of $dH_{c2}/dT$ for $P = 0$ and 20.4 kbar are about −8.0 and −6.7 T/K, respectively, suggesting that the values of $H_{c2}(0)$ could be extremely high.

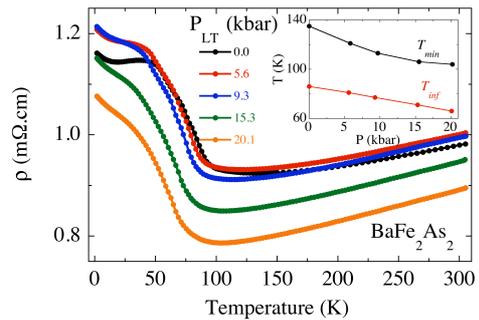

Fig. 3- (color online) Electrical resistivity vs temperature of BaFe$_2$As$_2$ in pressures up to 20 kbar. The inset shows the pressure dependence of the temperature of the minimum in $\rho(T)$, $T_{min}$, and $T_{inf}$, which reflect the onset of the crystallographic phase transition from high-T tetragonal to low-T orthorhombic.

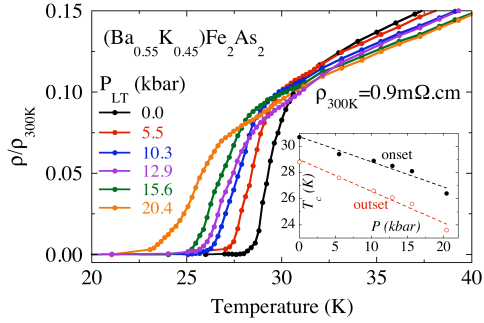

Fig. 4 – (color online) Normalized $\rho/\rho_{300K}$ vs $T$ for $(Ba_{0.55}K_{0.45})Fe_2As_2$ in pressures up to 20.4 kbar. The inset shows the pressure dependence of $T_{c,\,onset}$ (10%) and $T_{c,\,outset}$ (90%). The values of $dT_{c,\,onset}/dP$ and $dT_{c,\,outset}/dP$ are −0.21 K/kbar and −0.25 K/kbar, respectively.

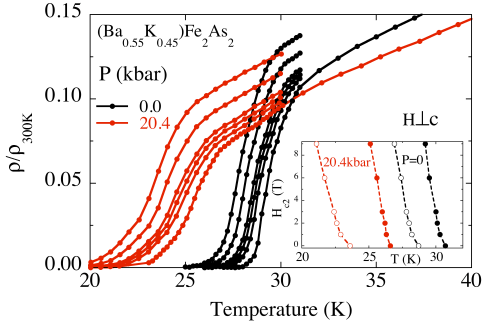

Fig. 5 – (color online) Normalized $\rho/\rho_{300K}$ vs $T$ of $(Ba_{0.55}K_{0.45})Fe_2As_2$ for $P = 0$ and 20.4 kbar, in $H = 0$, 1, 2, 3, 6, and 9 T. The inset shows $H_{c2,\,H\perp c}(T)$ for $P = 0$ and 20.4 kbar. Neglecting the low field upturns, the values of $dH_{c2}/dT$ for $P = 0$ and 20.4 kbar are approximately −8.0 and −6.7 T/K, respectively.

## 4. Conclusions

In conclusion, the effect of hydrostatic pressures up to 20 kbar is to suppress the structural transitions from high-T tetragonal (ThCr$_2$Si$_2$-type structure) to low-T orthorhombic in CaFe$_2$As$_2$ (totally) and BaFe$_2$As$_2$ (partially), at the approximate initial rates of −12 K/kbar and −1 K/kbar, respectively. Upon suppression of the orthorhombic transformation, CaFe$_2$As$_2$ displays superconductivity with onset near 12 K for $P > 2$ kbar. The further increase in pressure brings about another phase transformation, characterized by reduced scattering, and which is detrimental superconductivity. The correlation in temperature between the onset of the low-T-low-resistivity phase at $T_{s2}$ and the structural transformation from high-P-high-T tetragonal to high-P-low-T collapsed-tetragonal [Ref. 18] suggests that these could be one and the same. It remains to be seen whether fluctuations of the interactions responsible for the onset of either of the low-T phases within the SC pressure dome are in any way associated with the superconducting state. In light of the observation of superconductivity in BaFe$_2$As$_2$ in a dome extending from approximately 25-60 kbar, and in SrFe$_2$As$_2$ from ∼ 27-35 kbar, by means of magnetization measurements,[8] it would be important to the verify whether the behavior observed in CaFe$_2$As$_2$ is universal, i.e., the SC-dome being delimited at the low-P end by the suppression of the tetragonal to orthorhombic transformation, and at the high-P end by the transformation from tetragonal to collapsed-tetragonal.

K-doped (Ba$_{0.55}$K$_{0.45}$)Fe$_2$As$_2$ is tetragonal all the way down to low temperatures, and it displays superconductivity with $T_c$ neat 30 K. The effect of pressure is to lower $T_{c,\,onset}$ at the approximate rate of −0.21 K/kbar, and to broaden the SC transition slightly. The curves of $H_{c2}(T)$ shift down in temperature under pressure, and the magnitude of $dH_{c2}/dT$ drops slightly (from −8.0 to −6.7 T/K as $P$ is raised from 0 and 20.4 kbar), but it remains fairly high.

*Acknowledgments*

MST gratefully acknowledges support from the National Science Foundation under Grants No. DMR-0306165 and DMR-0805335. Work at the Ames Laboratory was supported by the US Department of Energy – Basic Energy Sciences under Contract No. DE-AC02-07CH11358.